\documentclass{osa-article}
\journal{oe}

\usepackage[output-decimal-marker={.},separate-uncertainty=true]{siunitx} 
\DeclareSIUnit\permille{\text{\textperthousand}}

\usepackage{xfrac}

\newcommand{\eq}[1]{\nolinebreak[5]Eq.~(\ref{#1})}			
\newcommand{\subscript}[1]{\nolinebreak[4]$\mathrm{_{#1}}$} 

\begin{document}

\title{The scaling potential of beam-splitter-based coherent beam combination}

\author{Michael Müller,\authormark{1,*} Christopher Aleshire,\authormark{1} Joachim Buldt,\authormark{1} Henning Stark,\authormark{1} Christian Grebing,\authormark{1,2} Arno Klenke,\authormark{1,3,$\dagger$} and Jens Limpert\authormark{1,2,3}}

\address{
\authormark{1}Friedrich Schiller University Jena, Institute of Applied Physics, Albert-Einstein-Straße 15, 07745 Jena, Germany\\
\authormark{2}Fraunhofer Institute for Applied Optics and Precision Engineering, Albert-Einstein-Straße 7, 07745 Jena, Germany \\
\authormark{3}Helmholtz-Institute Jena, Fröbelstieg 3, 07743 Jena, Germany 
}

\email{\authormark{*}michael.mm.mueller@uni-jena.de, \authormark{$\dagger$}a.klenke@gsi.de} 



\begin{abstract}
The impact of nonlinear refraction and residual absorption on the achievable peak- and average power in beam-splitter-based coherent beam combination is analyzed theoretically. While the peak power remains limited only by the aperture size, a fundamental average power limit is given by the thermo-optical and thermo-mechanical properties of the beam splitter material and its coatings. Based on our analysis, \SI{100}{kW} average power can be obtained with state-of-the-art optics at maintained high beam quality (M${}^\mathrm{2}\leq\num{1.1}$) and at only 2\% loss of combining efficiency. This result indicates that the power-scaling potential of today's beam-splitter-based coherent beam combination is far from being depleted. A potential scaling route to megawatt-level average power is discussed for optimized beam splitter geometry.
\end{abstract}



\section{Introduction}

The coherent combination of laser beams from many amplifiers allows producing in a single, much more powerful output beam, which circumvents the technical and fundamental-physical limitations of the amplifiers~\cite{brignon2013coherent}. This approach may enable visionary applications such as space-debris removal \cite{Soulard2014,Esmiller2014} and next-generation laser particle accelerators \cite{Leemans2011} that require diffraction-limited beam quality at \SI{100}{kW}-level average power, and in the latter case even ultrafast operation.

\begin{figure}[ht]
\centering\includegraphics[scale=1]{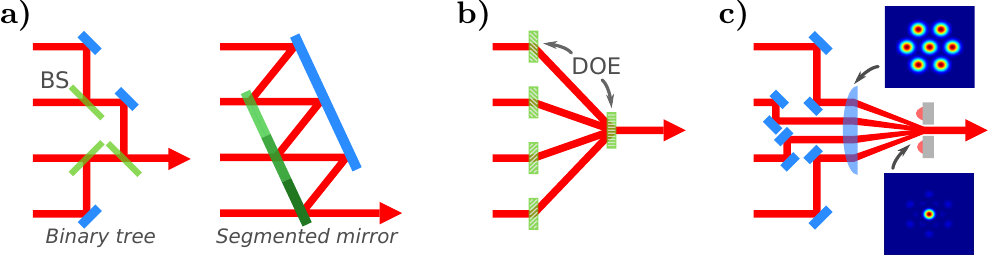}
\caption[]{Common CBC schemes: (a) Beam splitters (BS) in binary-tree and segmented-mirror arrangements, (b) diffractive optical elements (DOE) and (c) aperture tiling.}
\label{fig:CombiningMethods}
\end{figure}

There are many coherent beam combining techniques. The most common ones are depicted in Fig.~\ref{fig:CombiningMethods} and can be divided into two branches.
The first branch is filled-aperture combination in which the beams are overlapped collinearly using beam splitters (BS, \cite{Augst:07,uberna2010coherent}) or diffractive optical elements (DOE, \cite{Zhou2017,Redmond}). Ideally, these approaches allow for unity combining efficiency, defined as the ratio of the combined-beam power to the power of all input beams. In real systems, the efficiency typically is between 90\% and 95\% due to residual disparities of the input beams.
The second branch is aperture tiling in which an array of parallel beams is assembled and the diffraction into the far field poses the combination. The result is a central lobe containing the majority of the power and a pattern of side lobes, whose power cannot be diverted into the main feature. The power fraction in the central lobe is limited by the fill-factor of the input beam array, which for Gaussian beams at most is 68\%~\cite[p.~142]{brignon2013coherent} in the ideal case. Again, the real-case efficiency typically is lower at approximately 50\%, again due to the residual disparities of the input beams and practical constraints on the fill factor~\cite{Fsaifes2019}.

At the one hand, aperture tiling is power-scalable without limit as there is no physical beam combining element and the beam tiling introduces degrees of freedom for small-angle beam steering and beam shaping relevant to certain applications~\cite[Chap.~5.2]{brignon2013coherent}. However, its low efficiency is a drawback regarding high average-power operation.
At the other hand, filled-aperture schemes offer a high efficiency, but power scaling fundamentally is limited by residual absorption and nonlinear refraction in the beam combining elements.
These limits have been estimated for DOEs, concluding that MW-level average power and surface-damage-limited peak power can be obtained~\cite[Chap.~1.4]{brignon2013coherent}, but presently are unknown for beam-splitter-based combining. Therefore, these limits are derived in this contribution using a perturbative analysis to evaluate if beam-splitter-based combining can support the aforementioned visionary applications, too.

\section{Perturbative analysis of the two-beam interference}

\begin{figure}[h!]
\centering\includegraphics[scale=1]{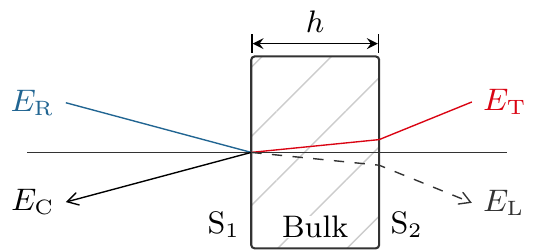}
\caption[]{Beam paths in a beam splitter of thickness $h$.}
\label{fig:geometry1}
\end{figure}

In all following analyses, two initially identical, pulsed laser beams are superposed on a beam splitter as depicted in Fig.~\ref{fig:geometry1}. One beam ($E_\mathrm{R}$) is reflected on the beam splitter front surface (S\subscript{1}), while the other beam ($E_\mathrm{T}$) is transmitted through the element passing both surfaces (S\subscript{1,2}) and the substrate material. Their interference creates a combined beam ($E_\mathrm{C}$) and a loss beam ($E_\mathrm{L}$). 

The interaction of the input beams with the surfaces and the bulk results in aberrations. They are described as phase deviations $\psi_i(\mathbf{x})$ of the beams' complex electric fields $E_i$, which are separated into real envelopes and phase terms as
\begin{equation}
E_i(\mathbf x) = A_i(\mathbf x) \exp\left(i \psi_i (\mathbf x)\right)
\label{eq:ansatzNEE}
\end{equation}
that depend on transverse space and time represented by the vector $\mathbf{x} = (x,y,t)$. 
This approach is valid, as the propagation lengths considered are short compared to the distances at which dispersive and diffractive effects significantly affect the envelopes. The combining efficiency $\eta$ is
\begin{align}
\eta 
=  \max _{\varphi \in [0,2\pi)}  \frac{\int | \sqrt{R}A_\mathrm{R}\exp\left( i \psi_\mathrm{R}\right) + \sqrt{1-R}A_\mathrm{T}\exp\left( i \psi_\mathrm{T}+i\varphi \right) |^2  dx^k}{ \int |A_\mathrm{R}|^2 + |A_\mathrm{T}|^2  dx^k} \text{,}
\label{eq:etac}
\end{align}
where $R$ is the power reflectivity of surface S\subscript{1}, $\varphi$ is the phase offset of the fields, and $dx^k$ denotes integration over the $k$ dimensions of $\mathbf{x}$. 
For small phase perturbations, the exponentials in Eq.~(\ref{eq:etac}) can be approximated to second order~\cite{Goodno2010a}. Then, the phase offset is chosen to compensate the phase difference $\psi(\mathbf{x})=\psi_\mathrm{R}(\mathbf{x})-\psi_\mathrm{T}(\mathbf{x})$ weighted by the optical power to solve the extremal problem. Lastly, the power ratio of the input beams is assumed to allow for a complete extinction of the loss beam in the ideal case. Then, the combining efficiency becomes
\begin{align}
\eta = 1 - R (1-R)  \mathrm{Var} [\psi(\mathbf{x})]
\label{eq:etavar}
\end{align}
that depends on the power-weighted variance of the beam's phase difference
\begin{align}
\mathrm{Var}[\psi(\mathbf{x})] & = \frac{\int \psi(\mathbf{x})^2 f(\mathbf{x}) dx^k}{\int f(\mathbf{x}) dx^k} - \left( \frac{ \int \psi(\mathbf{x}) f(\mathbf{x}) dx^k }{\int f(\mathbf{x}) dx^k} \right)^2 \text{,}
\label{eq:var}
\end{align} 
where $f(\mathbf{x})$ is the power weight. This approach gives a direct access to the combining efficiency in presence of phase aberrations and is applied to the cases of nonlinear refraction and residual absorption in the next sections.

\section{Efficiency degradation due to the Kerr effect}

In this section, the impact of nonlinear refraction and self-phase modulation on the combining efficiency is discussed for the case of Gaussian beams carrying Gaussian pulses. This requires finding the intensity-weighted phase variance of the transmitted beam only, as the reflected beam remains unaffected. Accordingly, \eq{eq:var} is evaluated using
\begin{equation*}
\psi(\mathbf{x}) = B_0 f(\mathbf{x}) \;\text{, }\; f(\mathbf{x}) = \exp({-x^2-y^2-t^2}) \;\text{, and }\; dx^3= dx  dy   dt \text{, } 
\end{equation*}
where $B_0= 2\pi \lambda^{-1} n_2 I_0 L_B$ is the peak nonlinear phase shift in the beam splitter substrate given by the center wavelength $\lambda$, the nonlinear refractive index $n_2$, the pulse peak intensity $I_0$, and the internal path length $L_B$, while $f(\mathbf{x})$ describes an amplitude-normalized, spatial and temporal Gaussian intensity envelope, and $dx^3$ is the differential volume in Cartesian coordinates. A short calculation yields
\begin{align*}
\mathrm{Var}[\psi(\mathbf{x})] = \left( \frac{1}{3\sqrt{3}} - \frac{1}{8}  \right) B_0^2 
\end{align*}
and the Kerr-degraded efficiency for a 50:50 power splitting ratio follows from \eq{eq:etavar} as
\begin{align}
\eta_\mathrm{K} = 1 - \frac{\mathrm{Var}[\psi(\mathbf{x})]}{4} \approx 1 - \frac{17}{1000} B_0^2 \text{.} \label{eq:K2}
\end{align}
Evaluating $\eta_\mathrm{K}$ for nanosecond-stretched pulses as emitted by chirped-pulse amplifiers reveals that optically induced damage rather than the Kerr effect is the prevailing limitation.
For example, considering \SI{1}{ns}-long pulses with a center wavelength of \SI{1}{\micro\meter} and a pulse energy of \SI{1.5}{J} in \SI{4}{mm}-diameter beams at close-to normal incidence on a \SI{5}{mm} thick fused silica beam splitter with a nonlinear refractive index of \SI{3.2e-20}{cm.^2\per W} results in an efficiency loss of only 0.1\%, while the combined-beam energy of \SI{3}{J} already is the surface damage limit~\cite{koechner2013solid}.

The nonlinear phase shift is relevant only when actual ultrashort pulses are involved. The combination of two \SI{1}{mJ}-pulses with \SI{200}{fs} pulse duration of typical ytterbium-based ultrafast fiber lasers in the same geometry would lead to 1\% efficiency loss, while being far from surface damage. In this case, an all-reflective solution using DOEs would be preferable as the Kerr effect is avoided altogether~\cite{Zhou2017}.

Eventually, both, the Kerr effect and the surface damage can be outrun as long as the beam diameter can be increased and the substrate thickness can be decreased. Hence, peak-power scaling with beam-splitter-based coherent beam combination is limited only by the technically achievable aperture size and diameter-thickness aspect ratio.

\section{Efficiency degradation due to residual absorption}

Now, the impact of thermal aberrations on the combining efficiency is analyzed. In general, the absorption of laser light in the bulk and on the surfaces of an optical element induces temperature gradients. The gradients causes a surface deformation and an inhomogeneous refractive index change that result in phase distortions of the interacting beams.

In the following, an analytic model of laser-induced heating is adapted to establish a link between the absorbed optical power and the combining efficiency. The influence of the beam splitter geometry and of the substrate material are discussed and the results are scaled to the combined-beam average power using measured and published absorption coefficients. 

\subsection{The aberration model}

There are exact analytic solutions of the internal temperature distribution $\Delta T_{\mathrm{S,B}}$ of a free-standing, cylindrical substrate heated by a Gaussian laser beam in axial symmetry for the cases of absorption on the surface ($\mathrm{S}$) or within the bulk ($\mathrm{B}$)~\cite[Eqs. (8) and (26)]{Hello1990}. The underlying model considers heat conduction within the bulk and the radiation off all surface, which requires a linearization of the Stefan-Boltzmann law that limits its validity to cases in which the surface temperature is not too different from the surrounding. This condition is targeted in coherent beam combining to minimize convection and always can be achieved with a sufficiently large beam diameter. Consequences of this approximation are that the absorbed power $P_a$ and the substrate peak temperature rise are linearly related and that the shape of the temperature distribution is determined by the geometry alone. This model originally developed for a vacuum environment is adapted to normal atmosphere by including a convection coefficient of \SI{10}{W\per{m^2K}}~\cite{cerbe2011technische}.

From the temperature distribution, the optical path length changes of the transmitted ($\mathrm{T}$) and reflected beam ($\mathrm{R}$) are calculated. For that, it is assumed that linear expansion takes place only in axial direction and that it is single-sided towards the partially reflecting surface yielding a worst-case estimate. Then, the optical path length changes are
\begin{align}
\mathrm{OPD}_\mathrm{R}(r) & = \left[- 2\alpha \right] \int_0^h  \Delta T_\mathrm{S,B}(r,z) dz \label{eq:OPDr} \\
\mathrm{OPD}_\mathrm{T}(r) & = \left[ (n-1)\alpha + \beta \right] \int_0^h  \Delta T_\mathrm{S,B}(r,z) dz\text{,} \label{eq:OPDt}
\end{align}
where $\alpha$ is the coefficient of linear expansion, $n$ is the refractive index of the medium at the surrounding temperature, $\beta$ is the temperature-dependent refractive index change, $h$ is the thickness of the beam splitter, and $r$ and $z$ are the radial and axial coordinates, respectively. The integral is carried out analytically and the difference of the obtained optical path length changes is converted to the phase error as 
\begin{align}
\psi(r) = 2\pi\lambda^{-1}\cdot \left[ \mathrm{OPD}_\mathrm{T}(r) - \mathrm{OPD}_\mathrm{R}(r) \right] \label{eq:psith}
\end{align}
in dependence of the laser wavelength $\lambda$. Finally, the thermally aberrated combining efficiency follows by numeric evaluation of Eqs.~(\ref{eq:var}) and (\ref{eq:etavar}) using the Gaussian-beam intensity weight $f(r) = \exp(-2r^2 w^{-2})$, where $w$ is the beam waist, and $dx^1 = r dr$.

\subsection{Example case: Intensity beam splitters}

The model is applied to a fused silica beam splitter with \SI{25}{mm} diameter and \SI{5}{mm} thickness for the combination of identical Gaussian beams with a waist of \SI{1.7}{mm}, which corresponds to the experimental setting of Ref.~\citenum{Muller2020}. The resulting temperature profiles for an absorption of \SI{110}{mW} average power are shown in the top panels of Fig.~\ref{fig:ThermCross} that are calculated with the MATLAB script attached in Supplement~1 and the material parameters of Ref.~\citenum{Hello1990a}. The radially resolved optical path length changes of the transmitted (red) and reflected (blue) beams are shown in the bottom panels as given by Eqs. (\ref{eq:OPDr}) and (\ref{eq:OPDt}). Their differences (dashed) are the relevant quantities for the efficiency calculation. The major contribution to the overall aberration is the thermally induced refractive index change ($\beta=\SI{8.6e-6}{K^{-1}}$) that is heavily affecting the transmitted beam, while the impact of the thermal expansion ($\alpha=\SI{.55e-6}{K^{-1}}$) is less relevant as apparent in the relatively small perturbation of the reflected beam. 

\begin{figure}[!h]
	\centering
	\includegraphics[width=\textwidth]{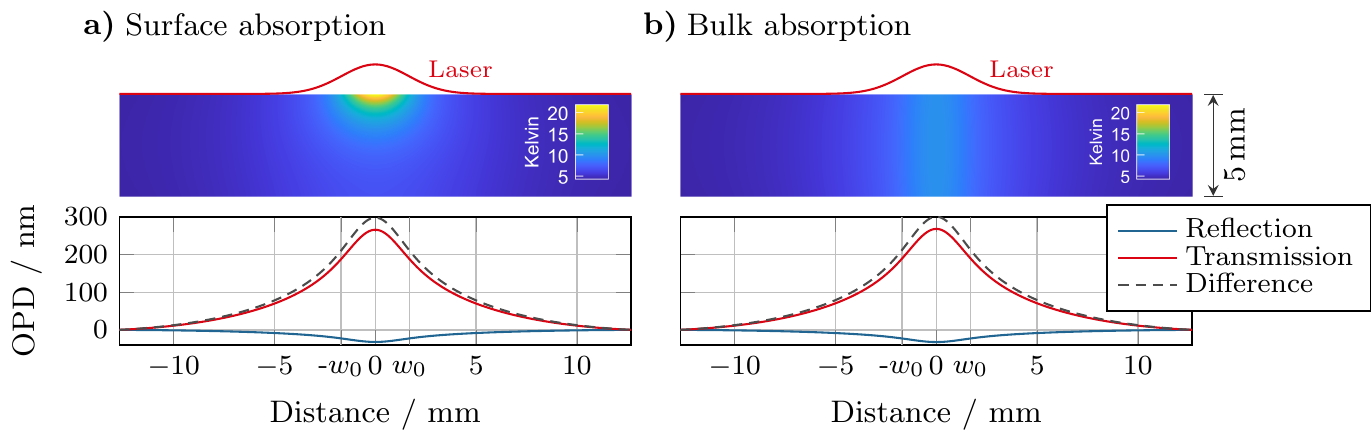}
	\caption[]{Cross-sectional temperature increase ($\Delta T$, top panel) and optical path length change ($\mathrm{OPD}$, bottom panel) for a cylindrical fused silica substrate heated by a Gaussian laser beam with with \SI{1.7}{mm} waist ($w_0$) incident from the top for absorption of \SI{110}{mW} on (a) the surface and in (b) the volume of the medium.}
	\label{fig:ThermCross}
\end{figure}

\begin{figure}[h]
	\centering
	\includegraphics[scale=.95]{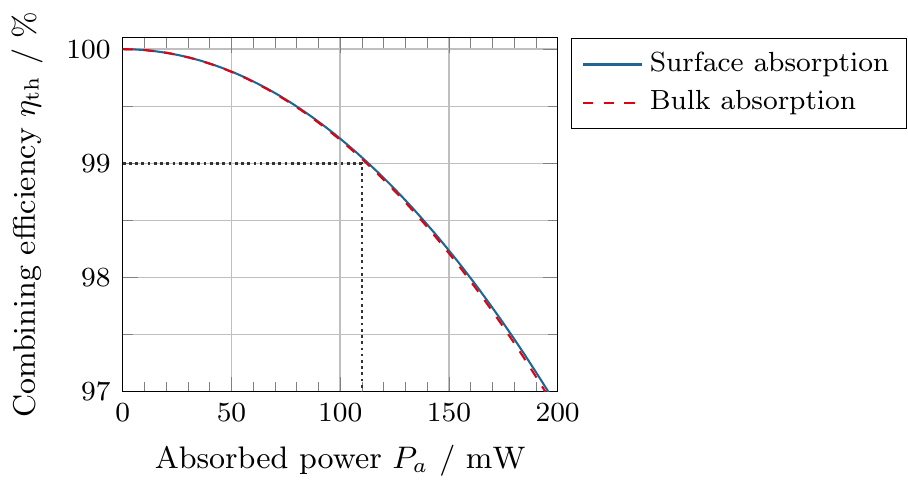}
	\caption[Combining efficiency versus absorbed beam power.]{Combining efficiency versus absorbed power in the example fused silica beam splitter with \SI{25}{mm} diameter and \SI{5}{mm} thickness for Gaussian beams with \SI{1.7}{mm} waist and \SI{1}{\micro m} wavelength. A critical absorbed power $P_\mathrm{cr}$ is defined at a 1\% efficiency loss, which is \SI{110}{mW} in this case (dotted line).}
	\label{fig:etatherm}
\end{figure}

The efficiency degradation shown in Fig.~\ref{fig:etatherm} follows presuming an equal power splitting ratio. In both modes of heating, the efficiency decreases quadratically for increasing absorbed power, which is a result from the linearized aberration model in the perturbative limit. Furthermore, there is virtually no difference regarding the two modes of heating, which is due to the heat load being dissipated from the interaction region primarily by conduction rather than re-emission within the beam area~\cite[Sec.\,A]{Winkler1991}. 
For the following, a critical absorbed power $P_\mathrm{cr}$ is defined that leads to a 1\% efficiency degradation (dotted line), which is about \SI{110}{mW} in fused silica in both modes of heating and to which the temperature distributions are shown in Fig.~\ref{fig:ThermCross}. In the following, it serves as figure of merit for the attainable combined-beam average power in a particular geometry.

\subsection{Beam splitter geometry}
\label{sec:geom}

\begin{figure}[h] 
\centering
\includegraphics[scale=.95]{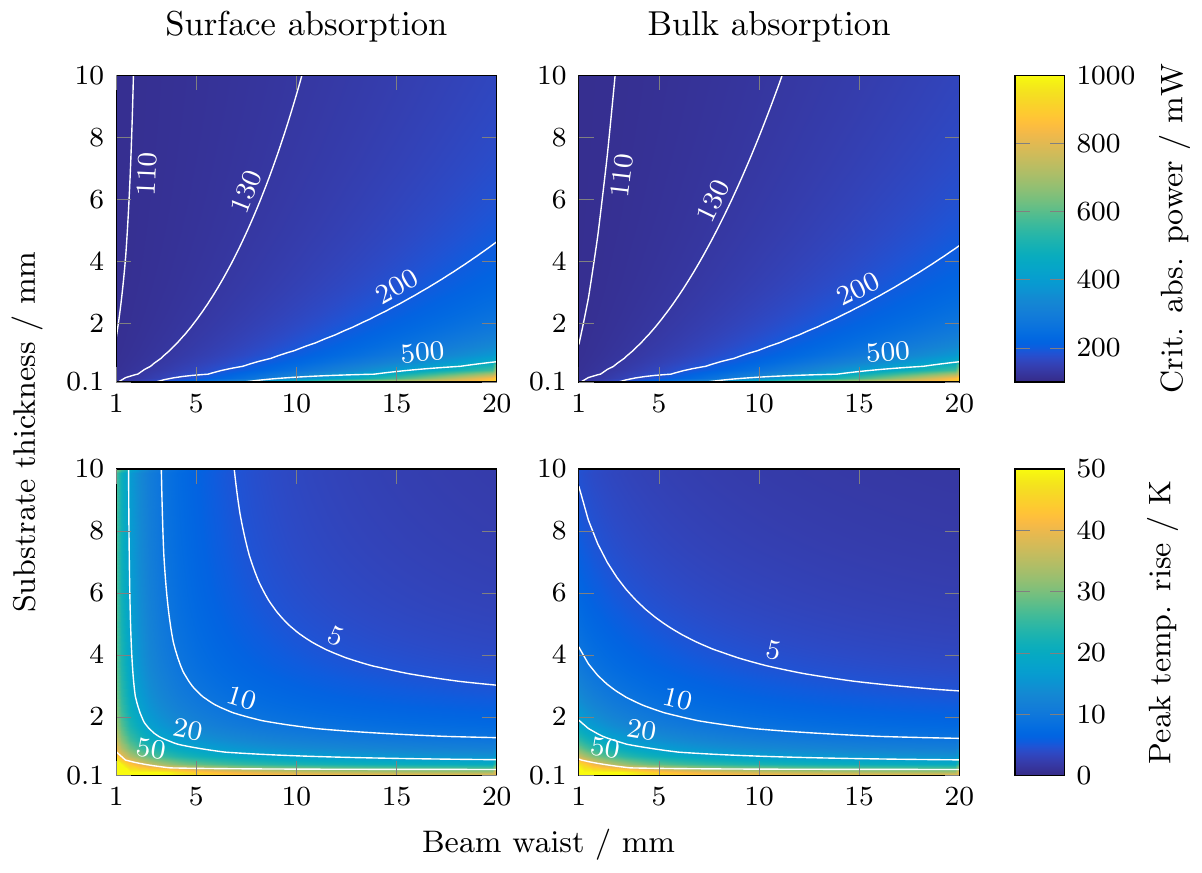}
\caption[]{Critical absorbed power (top panels) and peak temperature rise (bottom panels) for surface and bulk absorption in a fused silica beam splitter in dependence of the beam waist and the substrate thickness.}
\label{fig:EtaScan}
\end{figure}

Now, the influence of the beam splitter geometry on the attainable average power is investigated. For that, the critical absorbed power is computed for a range of substrate thicknesses and beam waists with the substrate radius is set to three times the largest occurring beam waist to avoid clipping. The results for the two modes of heating are shown in the top panels of Fig.~\ref{fig:EtaScan} and the corresponding peak temperature rise is shown in the bottom panels.

The critical absorbed power is a weak function of the geometry in most of parameter space with a value below \SI{200}{mW} in both modes of heating. This is due the heat load being dissipated primarily by radial conduction rather than direct re-emission from the interaction region, while the temperature gradient in axial direction is irrelevant.

A significant change of the critical absorbed power to \SI{1}{W}-level occurs for very thin substrates and very large beam waists at still tolerable surface temperatures. In this regime, the heat load is dissipated primarily by re-radiation and convection rather than conduction, which weakens the radial temperature gradient, in essence approaching the working principle of thin-disk lasers~\cite{Saraceno2019}.

This result has two implications for average-power scaling.
First, the critical absorbed power can be considered constant for typical 'thick' beam splitters with thickness-diameter ratios more than or equal to 1:5. Such ratio is typically required to achieve the mechanical stiffness supporting tenth-wave surface flatness guaranteeing an aberration-free reflection. For these beam splitters, the residual absorption poses a fixed, fundamental average-power limit that will be quantified shortly. 
Second, a significant scaling of the critical absorbed power could be achieved with a 'thin' beam splitter comprising a glass membrane stretched out on a frame, which may achieve sufficient flatness without the mechanical support of a 'thick' substrate. 
However, such a glass pellicle beam splitter has not been made, yet, asking for a feasibility study.
Hence, the following discussion focuses on the existing 'thick' beam splitters, for which the weak geometric dependence implies a large generality of the results.

\subsection{Scaling the results}
\label{sec:alpha}

The quantification of the average-power scaling limit requires a relation between the combined-beam average power $P_\mathrm{C}$ and the absorbed power $P_a$. This relation is established reconsidering the beam paths shown in Fig.~\ref{fig:geometry1}, which allows to equate
\begin{align}
P_a = \alpha_\mathrm{S_1}\left( P_\mathrm{R}+P_\mathrm{T} \right) + (\alpha_\mathrm{S_2} + \alpha_\mathrm{B} h) P_\mathrm{T} 
\overset{\text{(*)}}{\approx} \left( \alpha_\mathrm{S_1} + \zeta (\alpha_\mathrm{S_2} + \alpha_\mathrm{B} h) \right) P_\mathrm{C} := \alpha_\mathrm{eff} P_\mathrm{C}\text{,}
\label{eq:PB}
\end{align}
where $\alpha_\mathrm{S_{1,2}}$ and $\alpha_\mathrm{B}$ are the absorption coefficients of the surfaces and the bulk, $h$ is the path length in the bulk, $P_\mathrm{R,T}$ are the average powers of the reflected and transmitted beams, and $\zeta = P_\mathrm{T}(P_\mathrm{R}+P_\mathrm{T})^{-1}$ is the transmitted beam's share of the total input power.
At (*) it is assumed, that the sum of all input power is roughly equal to the combined-beam power, which is valid as long as the combining efficiency penalty is small. It allows to drop the loss beam from the discussion and to merge all absorption contributions into an effective absorption coefficient $\alpha_\mathrm{eff}$.
This approach falsely implies that the temperature distributions obtained in both modes of heating fulfill the superposition principle, but this error is small for small temperature changes.

The last missing piece are the absorption coefficients of the beam splitter's bulk and its coatings.
The considered bulk material is Suprasil~300, which is a high-purity, low-hydroxyl fused silica with an extremely low absorption coefficient $\alpha_\mathrm{B}$ of \SI{0.4\pm 0.3}{ppm\per cm} at \SI{1}{\micro\meter} wavelength~\cite{Heraeus2016a,Muhlig2008}.
Furthermore, anti-reflection (AR) coatings have been demonstrated with an absorption coefficient $\alpha_\mathrm{AR}$ of \SI{0.3}{ppm} at \SI{1}{\micro\meter} wavelength~\cite{Carpenter2012}. 
At the same time, high-reflection (HR) coatings feature a larger absorption coefficient $\alpha_\mathrm{HR}$ ranging from \SI{24}{ppm}~\cite{Steinlechner2012} down to \SI{4}{ppm}~\cite{Liu2017}, which is due to them being significantly thicker compared to anti-reflection coatings. Intermediate values are to be expected for the partially reflective (PR) beam splitter coatings, but no measured values are published so far. Thus, the thermographic image of the final beam combiner of the \SI{10.4}{kW} coherently-combined fiber laser depicted in Fig.~7 of Ref.~\citenum{Muller2020} is analyzed. 
The element is made of Suprasil~300, of which a \SI{5}{mm} thick uncoated sample was tested in a continuous wave, multimode ytterbium fiber laser at \SI{10}{kW} average power showing no detectable heating. 
Thus, the \SI{3}{K} temperature rise observed in Ref.~\citenum{Muller2020} can be assigned to coating absorption explicitly. It infers an absorbed power of \SI{15}{mW} from the combined beam at the \SI{3.3}{mm} beam diameter on the substrate, which can be read by linear scaling from Fig.~\ref{fig:ThermCross}(a). At the combined-beam power of \SI{13}{kW} before compression, this yields an absorption coefficient $\alpha_\mathrm{PR}$ of \SI{1.2}{ppm} for the PR coating, which falls right within the expected range.
%
%

Now, the combining efficiency shown in Fig.~\ref{fig:etatherm} can be related to the combined-beam power via the effective absorption coefficient
\begin{align}
\alpha_\mathrm{eff} = \frac{P_a}{P_\mathrm{C}} = \alpha_\mathrm{PR} + \zeta (\alpha_\mathrm{AR} + \alpha_\mathrm{B} h) = \SI{1.4}{ppm} \text{,}
\label{eq:PaIBS}
\end{align}
where $\alpha_\mathrm{PR}$, $\alpha_\mathrm{AR}$, and $\alpha_\mathrm{B}$ are as introduced above, $\zeta=0.5$, and $h=\SI{5}{mm}$. 

Lastly, the overall efficiency in a many-channel, symmetric binary tree as shown in Fig.~\ref{fig:CombiningMethods}(a) can be deduced from the final combination step. Considering Fig.~\ref{fig:etatherm}, the efficiency penalty reduces proportionally to the square of the absorbed power, which in turn reduces by $\sfrac{1}{2\eta}$ in each preceding combination step. Thus, a lower-bound estimate to the overall combining efficiency of a symmetric binary tree is
\begin{align}
\eta_{N} > \prod\limits_{n=0}^{N-1} \left( 1-\frac{\epsilon_0}{(2\eta_0)^{2n}} \right)
\quad \xrightarrow{N\rightarrow\infty} \quad \eta_{\infty}
\approx 1 - \left( 1 + \frac{1}{4\eta_0^2} \right) \epsilon_0  \text{,}
\label{eq:etainf}
\end{align}
where $\epsilon_0=\eta_0-1$ is the efficiency penalty of the final step. This product series converges rapidly, such that the linear contribution of the first two terms is sufficient in the perturbative limit.

\subsection{The average-power limit}
\begin{figure}[h]
\centering
\includegraphics[scale=1]{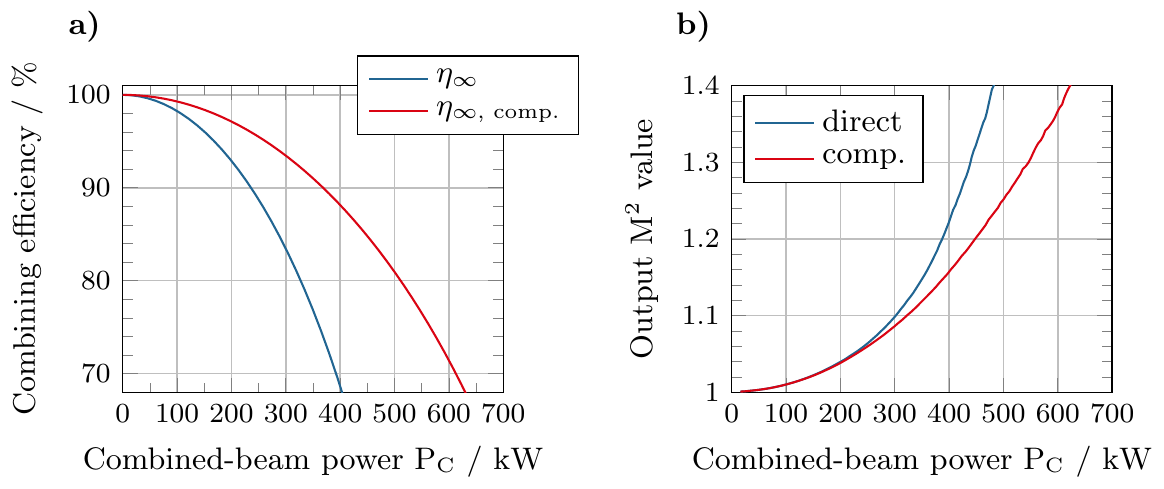}
	\caption[]{(a) Thermally aberrated combining efficiency of an infinite, symmetric beam splitter tree ($\eta_\infty$, blue). The red lines include a spheric compensation of the aberration. (b) The output  M${}^\mathrm{2}$ value for direct and spherically compensated beam combination.}
	\label{fig:etaIBS}
\end{figure}

Finally, the combining efficiency for an infinite binary tree comprising the aforementioned intensity beam splitters is obtained as shown in blue in Fig.~\ref{fig:etaIBS}(a) by scaling Fig.~\ref{fig:etatherm} with the effective absorption coefficient of Eq.~(\ref{eq:PaIBS}) and by scaling from the final-step efficiency to the overall efficiency using Eq.~(\ref{eq:etainf}). 
This result allows determining an average-power limit for 'thick' beam splitters in dependence of the acceptable combining efficiency loss as a first constraint. 
For the \SI{13}{kW} combined beam demonstrated in Ref.~\citenum{Muller2020}, the overall penalty was negligibly low at 0.1\%. Considering a 2\% loss tolerable would allow to increase the average power to \SI{100}{kW}. Beyond that, the efficiency declines rapidly until 68\% efficiency is reached at \SI{400}{kW} average power. From this value on, tiled-aperture combining would be more efficient for the considered Gaussian beams.

A second constraint is the peak temperature of the beam splitters. It is linearly related to the absorbed power and approximately linearly related to the beam radius~\cite[Eq.~4]{Winkler1991}. For example, the \SI{22}{K} gradient shown in Fig.~\ref{fig:ThermCross}(a) is certainly tolerable and corresponds to a combined-beam power of \SI{82}{kW}. Maintaining this temperature rise implies a beam waist of \SI{8.3}{mm} would be required for the \SI{400}{kW} operation point, which could be realized on a \SI{2}{"}-diameter substrate.

The efficiency under thermal load could be increased if additional complexity is allowed. For example, a spherically concave beam splitter backside could partially alleviate the wavefront mismatch considering the dashed curves shown Fig.~\ref{fig:ThermCross}. Numeric optimization of this curvature results in an increased efficiency shown in red in Fig.~\ref{fig:etaIBS}(a). Then, even \SI{500}{kW} average power could be achieved still at 80\% combining efficiency. Also, higher order phase aberration could be compensated with aspheric surfaces at a corresponding increase of either the attainable average power or the efficiency. 

A third constraint, besides the combining efficiency and the surface temperature, is the combined-beam quality that is relevant for many applications. Thus, the combined-beam output M${}^\mathrm{2}$ value is shown in Fig.~\ref{fig:etaIBS}(b) for the cases of direct and spherically compensated combination. These values are computed numerically by applying the accumulated phase distortion of 10 preceding combination steps of the binary tree (= 1024 combined channels) to an ideal Gaussian beam. The M${}^\mathrm{2}$ value then is found by fitting a parabola to the beam caustic obtained by numeric beam propagation through a focus~\cite{iso2005lasers}. This approach neglects the beam propagation between subsequent beam splitters, which is valid as these distances are short compared to the typical Rayleigh length. The result shows that, in either case, the M${}^\mathrm{2}$ value degrades from initially \num{1} to \num{1.1} for a combined-beam average power of \SI{300}{kW}, which can be considered a limit for applications demanding diffraction-limited beam quality. Adaptive beam shaping might leverage the phase distortion of the output beam such that further scaling would be possible~\cite{Wattellier2004}.

Reconsidering \eq{eq:PaIBS}, the largest share of the absorption in the beam combiner stems from the partially reflective coating, which in Ref.~\citenum{Muller2020} already was optimized for least absorption in terms of the coating materials and the general coating process, but even more aspects influence the residual absorption. For example, superpolishing~\cite{Juskevicius2013,Kamimura2004}, removal of damage precursors during the coating process~\cite{Thimotheus2020}, and the minute minimization of the hydroxyl content in the coating layers might allow to reduce this absorption contribution even further and, with that, to increase the combined-beam average power.

\paragraph{Pellicle-like beam splitters}
The abscissas of Fig.~\ref{fig:etaIBS} can be rescaled from the 'thick' beam splitters to the 'thin' pellicle geometry discussed in Sec.~\ref{sec:geom} by the ratio of their critical absorbed powers, which is a factor $\sfrac{\SI{1000}{mW}}{\SI{110}{mW}}=9$. This implies an average power in excess of \SI{1}{MW} could be reached, if the pellicle geometry is technically feasible for centimeter-sized beam radii.

\paragraph{Thin-film polarizers}

Other coherent beam combining systems rely on polarization beam combination using thin-film polarizers (TFP)~\cite{Stark2021}. For those, the axially symmetric model can be applied despite the oblique incidence, since the geometric dependence of the thermal aberration is weak, allowing to rescale the previous results accordingly. In particular, considering a \SI{5}{mm}-thick Brewster-type TFPs made of Suprasil~300 with a TFP coating that presumably is as good as an optimized HR coating~\cite{Liu2017}, yields an effective absorption coefficient
\begin{align}
\alpha_\mathrm{eff,TFP} = \alpha_\mathrm{HR} + \zeta \alpha_\mathrm{B} h = \SI{4}{ppm} + 0.5\cdot \SI{0.5}{cm}\cdot \SI{0.4}{ppm\,cm^{-1}} = \SI{4.1}{ppm} \text{.}
\label{eq:PaTFP}
\end{align}
As this coefficient is about 3 times larger than in the previous case of the intensity beam splitter, the abscissas in Fig.~\ref{fig:etaIBS} have to be divided by 3. Thus, the achievable average power for 'thick' TFPs is about \SI{33}{kW} at an efficiency penalty of 2\%. At the same time, a correction of the spherical aberration term by using concave backsides of the TFPs is complicated due to the oblique incidence, i.e. the red curves of Fig.~\ref{fig:etaIBS} do not apply.

\paragraph{Wavelength scaling}

Most of today's coherent beam combining systems are based on ytterbium-doped amplifiers because they are developed to their fundamental limits already. Thus, all discussions herein are focused on \SI{1}{\micro\meter} wavelength, but the conversion to other wavelengths is straightforward. 
For example, considering Eq.~(\ref{eq:psith}) for thulium-based \SI{2}{\micro\meter} lasers and assuming that the absorption coefficients remain unchanged implies that the same optical path length error yields only half the wavefront error. Then, the square-dependence of Eq.~(\ref{eq:etavar}) yields only a quarter of the efficiency penalty for the same absorbed power in comparison to \SI{1}{\micro\meter} wavelength. Conversely, a fourfold higher average power is attainable at the same efficiency penalty.

\subsection{Alternate substrate materials}
The so far considered dry fused silica is not necessarily the best substrate material for a beam-combining element. Highly heat-conducting sapphire or synthetic diamond could offer a power-scaling advantage considering their favorable ratios of heat conductivity to thermal expansion and thermal refractive index change.

To evaluate this possibility, the aberration model is adapted with the published material parameters for sapphire~\cite{Patel1999,weber2002handbook}, which yields a critical absorbed power of \SI{800}{mW} in the example geometry. Compared to fused silica, this approximately 7-fold increase stems from the significantly larger thermal conductivity spreading out the heat more evenly at a similar temperature-dependent refractive index change, despite the one order of magnitude larger thermal expansion.
At the same time, the bulk absorption coefficient $\alpha_\mathrm{Sapp}$ also is two orders of magnitude larger compared to dry fused silica at about \SI{40}{ppm\per\centi\meter} in high purity material~\cite{GTAT}. Now, swapping the Suprasil~300 for sapphire in \eq{eq:PaIBS} yields an effective absorption coefficient of $\alpha_{\mathrm{eff,Sapp}}=\num{11e-6}$, which is 8-times larger than before and diminishes the 7-fold-increase of the critical absorbed power. Thus, no net average-power increase is enabled for the case of intensity beam splitters. This is different in the case of thin-film polarizers (Eq.~(\ref{eq:PaTFP})), as the effective absorption coefficient increases only by a factor of 4, which allows a two-fold increase in the average power. 

For synthetic diamond, the critical absorbed power is about three orders of magnitude larger than in fused silica due to its enormous thermal conductivity~\cite{Slack1975} while the remaining thermo-mechanical and thermo-optical properties are comparable~\cite{Yurov2017}. However, the bulk absorption is also three orders of magnitude larger compared to fused silica~\cite{Webster2015}, again resulting in no significant benefit for either type of combining architecture. 

Thus, dry fused silica remains the best substrate material for intensity beam splitters, while sapphire may offer a small benefit in the case of thin-film polarizers.

\section{Summary}

Coherent beam combination allows for power scaling beyond the fundamental limits of laser amplifiers. Implementations based on binary beam splitters offer a low cost and high efficiency, but eventually residual absorption and nonlinear refraction induce efficiency-impairing aberrations. The influence of these aberrations on the power scalability was quantified in a perturbative analysis. 

The peak power is limited either by optically induced damage of the surface in the nanosecond-stretched-pulse regime or by phase aberration through nonlinear refraction when actual ultrashort pulses are considered. An increasing aperture size mitigates both effects.


The average power is limited by thermal aberration due to residual absorption. This effect was analyzed for surface and bulk absorption using existing analytic solutions of the substrate-internal temperature distribution. 
State-of-the-art intensity beam splitters with \SI{1.4}{ppm} residual absorption support an average power of \SI{100}{kW} at 98\% combining efficiency in a binary-tree scheme without noticeable loss of beam quality. Concave beam splitter backsides would allow for an additional 1.5-fold power increase. Considering an M${}^\mathrm{2}$ degradation of 0.1 the limit towards applications that require a diffraction-limited beam quality allows for up to \SI{300}{kW} average power at still above 90\% combining efficiency and at a beam radius of about \SI{1}{cm} to maintain tolerable temperatures.
%
%
Pellicle-like beam splitters are identified as the thermally most resilient geometry that minimizes the efficiency-degrading radial temperature gradient. They could enable megawatt-level average output power given they are technically feasible. In addition, a further reduction of the residual absorption, foremost in the coatings, would improve on the achievable average power. Sapphire and diamond were considered as alternate substrate materials, but in both cases the advantage of their smaller thermal aberration per absorbed power is abated by their larger bulk absorption such that no significant advantage is expected using either material.

In the end, the fundamental difference of beam splitters for coherent beam combination in comparison to reflective diffractive optical elements or aperture tiling is that always one beam has to pass through the substrate material. 
It is the major cause of the efficiency impairing aberrations, such that the attainable peak or average power is limited by the technical ability of making thin, large-aperture beam splitters of high surface quality. 
However, the state-of-the-art beam splitters already support power scaling beyond \SI{100}{kW}-level average power and Joule-level pulse energy, which already suffices to enable even the most demanding applications envisioned today.

\begin{backmatter}
\bmsection{Funding} 
Fraunhofer-Gesellschaft (Cluster of Excellence “Advanced Photon Sources”); European Research Council “SALT” (835306).


\bmsection{Disclosures} 
The authors declare no conflicts of interest.

\bmsection{Data Availability}
Data underlying the results presented in this paper are not publicly available at this time but may be obtained from the authors upon reasonable request.
\end{backmatter}

\bibliographystyle{osajnl}
\bibliography{library}{}

\end{document}